\pdfoutput=1
\documentclass[a4paper,american,citeautoscript,floatfix,pdftex,superscriptaddress,twoside,%
aps,pra,%
article%
]{revtex4-2}%

\usepackage[top=2.5cm,bottom=2.5cm,left=2.5cm,right=2.5cm]{geometry}
\usepackage{graphicx}
\usepackage{float} 
\usepackage{amsmath}
\usepackage{subfig}
\usepackage{amssymb}
\usepackage{hyperref, hypernat}
\usepackage{beitmacros}

\newcommand{\beit}{\affiliation{BEIT sp. z o o., Mogilska 43, 31-545 Kraków, Poland}}%
\newcommand{\beitemail}{\email[]{emil@beit.tech}}%
\newcommand{\beitweb}{\homepage{https://www.beit.tech}}%

\graphicspath{{./}}

\begin{document}

\title{Utilizing redundancies in Qubit Hilbert Space to reduce entangling gate counts in the Unitary Vibrational Coupled-Cluster Method}
\author{Michal Szczepanik and Emil Zak}\beit\beitemail\beitweb

\begin{abstract}
	We present a new method for state preparation using the Unitary Vibrational Coupled-Cluster (UVCC) technique. Our approach utilizes redundancies in the Hilbert space in the direct mapping of vibrational modes into qubits. By eliminating half of the qubit controls required in the Trotterized UVCC ansatz, our method achieves up to a 50\% theoretical reduction in the entangling gate count compared to other methods and up to a 28\% reduction compared practically useful approaches.  
	This improvement enhances the fidelity of UVCC state preparation, enabling more efficient and earlier implementation of complex quantum vibrational structure calculations on near-term quantum devices. We experimentally demonstrate our method on Quantinuum’s H1-1 quantum hardware, achieving significantly higher fidelities for 6- and 8-qubit systems compared to existing implementations. For fault-tolerant architectures, eliminating half of the control qubits in multi-controlled rotations incurs an additional Toffoli gate overhead elsewhere in the circuit. Thus, the overall performance gain depends on the specific decomposition method used for multi-controlled gates.
\end{abstract}

\maketitle

\section{Introduction}
Born-Oppenheimer vibrational energies are obtained as solutions to the Schroedinger equation for the motion of nuclei on a potential energy surface (PES) \cite{Bunker1999}. Achieving high-accuracy vibrational energy levels, at 1\invcm or better, requires variational or pseudospectral methods and a high-quality PES \cite{Bowman2008, Carrington2017}. Both methods transform quantum mechanical differential equations into a matrix eigenvalue problem, the size of which scales exponentially with the number of vibrational modes.  
For this reason, despite their relevance to industry and fundamental science, high-quality nuclear motion calculations have been limited to a few atoms~\cite{Avila2019,Carrington2017}. Quantum computers are expected to alleviate this limitation~\cite{Sawaya2021,Ollitrault2020}.

The development of computational methods, whether for quantum or classical devices, that address the exponential curse of dimensionality in rotational-vibrational calculations is important, given their applications in molecular spectroscopy (reference spectra for experiments~\cite{Barber2006}), astrophysics (e.g., interstellar medium composition~\cite{Motiyenko2012}), exoplanet research (e.g., atmospheric retrirevals~\cite{Madhusudhan2018,Tennyson2024}), material design (e.g., simulating phonons in crystals), and nanoelectronics (e.g., designing molecular junctions~\cite{Erpenbeck2016,Ke2023} and modelling vibrational relaxation~\cite{Meyer2006}). Notably, among plurality of other uses, vibrational calculations aid in the design and operations of trapped-ion~\cite{Chen2021,Hucul2014,Bruzewicz2019,strohm2024} and superconducting~\cite{Kitzman2023,Rosen2019} quantum computing architectures.

Despite the unfavourable scaling, the variational approach carries the advantage of controlling error by adjusting the basis set size and quality. In this spirit, near-term quantum computing devices adopted the Variational Quantum Eigensolver (VQE) for vibrational calculations~\cite{Peruzzo2014, Romero2018, Tilly2022}. In VQE, a trial state is prepared by executing a quantum computing circuit, parametrised by variational coefficients that are optimised on a classical computer. However, the classical optimization process and the need for sampling multi-dimensional distributions limit the scalability of VQE~\cite{Tilly2022}.  An alternative algorithm suited for fault-tolerant quantum architectures is quantum phase estimation (QPE)~\cite{Abrams1997,Abrams1999,AspuruGuzik2005}, which returns energy levels for a given trial state and a unitary embedding of the Hamiltonian~\cite{Chakraborty2019,Gilyn2019,Low2019}. 

In both VQE and QPE, the quality of initial trial state preparation is essential for accurate results~\cite{DCunha2024,fomichev2023}. In VQE, the space spanned by the ansatz sets a lower bound on the calculated energy levels, while in QPE, the ansatz's overlap with the target eigenstate determines the likelihood of correctly measuring the corresponding eigenvalue.
Typically, the larger the basis set representing the trial state vectors, the better the accuracy of the energy levels. For this reason, excitation-based formats for the variational ansatz have gained popularity. Classical computing techniques including vibrational configuration interaction (VCI)~\cite{Mathea2021} and vibrational coupled-cluster (VCC)~\cite{Christiansen2004} allow control over the number of basis functions, through the choice of excitation level. 
For this reason, a quantum computing counterpart to VCC, the unitary vibrational coupled-cluster (UVCC) method, has been employed for state preparation in VQE~\cite{Mcardle2019,Sawaya2020}, and it is also applicable to QPE~\cite{Lin2020,halder2021,DCunha2024}. While the quality of the UVCC ansatz can be refined by increasing excitation levels or by enhancing the quality of the reference basis, 
such improvements lead to rapid increases in the computational complexity, particularly in terms of two-qubit CNOT gate count~\cite{Sawaya2020,Ollitrault2020}. Quality of the basis set can be improved through solving reduced-dimensionality Schroedinger equations with effective Hamiltonians, as in the multimode procedure, although this sacrifices the Hamiltonian's sparsity~\cite{Bowman2003,Bowman2008,Carrington2017}. 

Upon encoding the UVCC operator into a qubit circuit, irrespective to encoding type, one faces artifacts that affect the efficiency of the quantum computing procedure, when no improvements are made~\cite{Sawaya2020}. For example, in unary encoding each vibrational basis state $\ket{n}$ is represented by a qubit in state $\ket{1}$ of an appropriate index, leaving a grossly large fraction of the Hilbert space nonphysical. This spurious part of the Hilbert space does not improve the solution. On the other hand, unary encoding is advantageous to compact binary encodings due to a lower number of entangling gates~\cite{Sawaya2020}. 

In this work, we propose a method for constructing the UVCC ansatz that requires fewer CNOT gates compared to other approaches. Our proposed procedure leverages on the awareness of a computational waste carried along the quantum computation using unary encoding. Specifically, we observe that a considerable portion of the qubit Hilbert space does not contribute to the solution. By eliminating control over spurious states, our method achieves up to a 50\% reduction in CNOT gates compared to methods reported in previous studies \cite{Mcardle2019, Ollitrault2020,Ltstedt2022}, including techniques based on Givens rotations~\cite{Delgado2021,Arrazola2022}. This advantage is apparent for both low-excitation UVCC cases and in the asymptotic limit, remaining largely insensitive to the decomposition method for multi-controlled NOT gates. However, compared to techniques that employ ancillary qubits to decompose multi-controlled NOT gates into CNOTs, such as the recent method of Khattar and Gidney~\cite{khattar2024riseconditionallycleanancillae}, the relative advantage of our approach is typically less pronounced, leveling off at around 28\%.

In the context of fault-tolerant quantum computing, our approach reduces the number of control qubits needed in controlled rotations within the UVCC circuit by half, though this requires an additional $\mathcal{O}(m)$ Toffoli gate overhead, where $m$ represents the vibrational excitation level. Integrating our method with ancilla-assisted decomposition schemes for multi-controlled rotations offers limited overall reductions in Toffoli gate counts, depending on the chosen decomposition strategy. While optimizing Toffoli gate counts is not the focus of this work, the information-theoretic reductions presented here provide opportunities for such optimizations.

\section{The Unitary Vibrational Coupled-Cluster ansatz}
\label{sec:UVCC}
    In vibrational structure calculations, a fixed number of modes $M$ is associated with a set of internal coordinates $\{Q_i\}$ of the system. For simplicity, we choose the harmonic oscillator basis, since it allows for a straightforward mapping to qubits~\cite{Sawaya2020}. However, the following method can be generalized to other basis sets, as described in \cite{Ollitrault2020}. In \textit{direct} or \textit{unary} mapping, $d$ lowest eigenstates $\ket{n}$ of a 1-dimensional harmonic oscillator are encoded as follows
    \begin{equation}
        \ket{n} = \bigotimes_{j=0}^{n-1}\ket{0}_j\ket{1}_n\bigotimes_{j=n+1}^{d-1}\ket{0}_j.
    \end{equation}
    Despite the inefficient use of qubits, this mapping generates shallower circuits than any compact mapping (binary encoding).
    The Unitary Vibrational Coupled-Cluster ansatz is obtained by applying a cluster operator to the reference state~\cite{Mcardle2019}
    \begin{equation}
        \ket{\Psi(\boldsymbol{\theta})} = \exp(\hat{T}-\hat{T}^\dag)\ket{\Psi_0}
        \label{eq:UVCC_ansatz}
    \end{equation}
    where $\boldsymbol{\theta}$ denotes the set of parameters and $\hat{T}$ is a sum of excitation operators:
    \begin{equation}
        \hat{T} = \hat{T}_1+\hat{T}_2 + ...
    \end{equation}
    and
    \begin{equation}
        \hat{T}_1 = \sum_{i=1}^M\sum_{n_i,m_j=0}^{d-1}\theta_{n_i,m_i}\ket{n_i}\bra{m_i}
    \end{equation}
    \begin{equation}
        \hat{T}_2 = \sum_{i<j}^M\sum_{n_i,m_i,k_j,l_j=0}^{d-1}\theta_{n_i,m_i,k_j,l_j}\ket{n_i}\bra{m_i}\otimes\ket{k_j}\bra{l_j}
    \end{equation}
    \begin{equation*}
        \vdots
    \end{equation*}
    Upon Trotterization with the direct mapping the cluster operator in eq.~\ref{eq:UVCC_ansatz} can be cast as a product of $m$-excitation unitaries acting on appropriate qubits~\cite{Mcardle2019,Ollitrault2020}. These unitaries, up to the reordering of qubits, all have the following form
    \begin{equation}
        \hat{U}_m(\theta) = \exp{\left[\theta \left(\ket{e_m}\bra{g_m}-\ket{g_m}\bra{e_m}\right)\right]}
        \label{eq:exc_unitary}
    \end{equation}
    where,
    \begin{align}
        \ket{g_m} = \bigotimes_{i=0}^{m-1}\ket{01}_i &&\ket{e_m} = \bigotimes_{i=0}^{m-1}\ket{10}_i.
    \end{align}
 
    In a standard approach, utilized for example in \cite{Mcardle2019,Ollitrault2020}, the $U_m(\theta)$ gates are expressed in terms of Pauli operators using the relation $\ket{0}\bra{1} = \frac{1}{2}(X+iY)$, where $X$ and $Y$ are single-qubit Pauli-X and Pauli-Y operators, respectively. Another Trotterization is then applied, giving $2^{(2m-1)}$ gates per Pauli operator, which can be decomposed using $2(2m-1)$ CNOT gates. Therefore, this method requires $4^m(2m-1)$ 2-qubit gates to implement the $m$-excitation unitary. We call this method \textit{exponential decomposition} from here on.

    More efficient approaches rely on the fact that unitaries defined in eq.~\ref{eq:exc_unitary} perform a 2-dimensional rotation in the space spanned by $\ket{g_m}$ and $\ket{e_m}$, while leaving other states unchanged:
    \begin{equation}
        \hat{U}_m(\theta)\ket{g_m} = \cos\theta\ket{g_m}+\sin\theta\ket{e_m}
        \label{eq:rot1}
    \end{equation}

    \begin{equation}
        \hat{U}_m(\theta)\ket{e_m} = -\sin\theta\ket{g_m}+\cos\theta\ket{e_m}
        \label{eq:rot2}
   \end{equation}
    and
    \begin{equation}
        \hat{U}_m(\theta)\ket{l} = \ket{l}
        \label{eq:rot3}
    \end{equation}
    for $\bra{l}g_m\rangle = \bra{l} e_m\rangle = 0$.
    The $\hat{U}_m(\theta)$ is then expressed as~\cite{yordanov2020,Anselmetti_2021,Nam2020}
    \begin{equation}
        \hat{U}_m(\theta) = \hat{\mathcal{U}}_m^\dag RY\left(\theta, \{q_1..q_{2m-1}\}, q_0\right)\hat{\mathcal{U}}_m
       \label{eq:U_decomp}
    \end{equation}
    where $\mathcal{U}_m$ transforms states $\ket{g_m}$,$\ket{e_m}$ so that they only differ in the first qubit and $RY\left(\theta, \{q_1..q_{2m-1}\}, q_0\right)$ is a Y-rotation on $q_0$ controlled by qubits ${\{q_1..q_{2m-1}\}}$. With such an approach the 2-qubit gate cost for the double excitation unitary ($m=2$) reduces from 48 for the exponential decomposition to 13~\cite{Nam2020}. The method of ref.~\cite{Nam2020} uses 3 CNOT gates to represent $\mathcal{U}_2$. For a general excitation level $m$ the  method requires $(2m-1)$ CNOT gates. The total cost is thus $2(2m-1)$ CNOT gates plus the implementation of the $(2m-1)$-controlled Y-rotation.

     We propose a method which, for the case of molecular vibrational calculations within the direct mapping, implements the $m$-excitation unitary with $(8m-6)$ CNOT gates and a single $m$-controlled Y-rotation. We note that in the direct mapping, only a small subspace of the qubit Hilbert space is utilised. Therefore, it is sufficient for a decomposed gate to transform correctly states in this used subspace only. However, this means that the method is less general than the methods of refs.~\cite{Arrazola2022,yordanov2020,Anselmetti_2021,Nam2020}. For example, the present method cannot be straightforwardly extended to state preparation with the unitary coupled cluster method for the electronic structure problem.

\section{Procedure}
\label{sec:procedure}
\begin{figure}[h]
    \centering
    \includegraphics[width=0.6\textwidth]{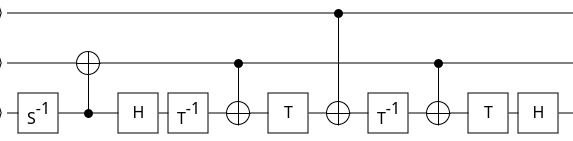}
    \caption{Decomposition of the $\hat{\mathcal{S}}_j$ gate defined in eq.~\ref{eq:Sj-gate} into single qubit gates $S,T,H$ and CNOTs.}
    \label{fig:H2}
\end{figure}

\begin{figure}[h]
    \centering
    \includegraphics[width=0.5\textwidth]{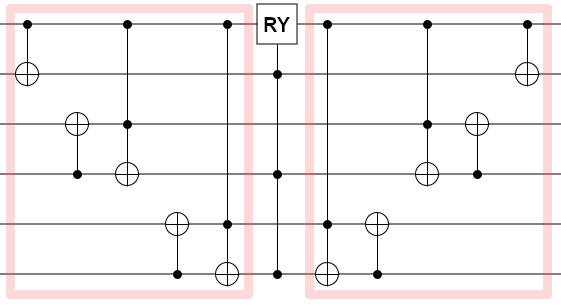}
    \captionsetup{justification=raggedright, singlelinecheck=false}
    \caption{Quantum circuit representing the $m$-excitation unitary defined in eq.~\ref{eq:vib_U} for $m=3$. Qubits are numbered from 0 to 5, top to bottom. The left red box corresponds to $\hat{\mathcal{U}}_m$ and the right red box corresponds to $\hat{\mathcal{U}}^\dag_m$.}
    \label{fig:example}
\end{figure}

In our method, the $m$-excitation unitary defined in eq.~\ref{eq:exc_unitary} is represented as
    \begin{equation}
        \hat{U}_m(\theta) = \hat{\mathcal{U}}_m^\dag RY\left(\theta, \{q_{1},q_{3}..,q_{2m-1}\},q_0\right)\hat{\mathcal{U}}_m
        \label{eq:vib_U}
    \end{equation}
	such that the Y-rotation about angle $\theta$ is controlled on every other qubit, different to the decomposition of eq.~\ref{eq:U_decomp}, where control is on all qubits.
	We define the $\hat{\mathcal{U}}_m$ gate as follows
    \begin{equation}\label{transf_basis}
        \hat{\mathcal{U}}_m = CX(q_0,q_1)\prod_{j=1}^{m-1} \hat{\mathcal{S}}_j
    \end{equation}
    where $\hat{\mathcal{S}}_j$ is a 3-qubit gate applied to the qubits of the $j$-th mode and the first qubit:
    \begin{equation}
        \hat{\mathcal{S}}_j = CX(q_{2j+1},q_{2j})RTOF(\{q_0,q_{2j}\}, q_{2j+1})
        \label{eq:Sj-gate}
    \end{equation}
    The $RTOF$ gate is the relative phase Toffoli gate, which can be implemented with 3 CNOT gates \cite{Maslov2015}. A circuit implementing $\hat{\mathcal{S}}_j$ is shown in Figure~\ref{fig:H2}.
    \noindent
    The $\hat{\mathcal{S}}_j$ gate transforms the appropriate states in the following way:
    \begin{align}
        \hat{\mathcal{S}}_j\ket{00}_j\ket{0}_0 = \ket{00}_j\ket{0}_0  && \hat{\mathcal{S}}_j\ket{00}_j\ket{1}_0=\ket{00}_j\ket{1}_0
    \end{align}
    \begin{align}
        \hat{\mathcal{S}}_j\ket{01}_j\ket{0}_0 = \ket{01}_j\ket{0}_0 && \hat{\mathcal{S}}_j\ket{01}_j\ket{1}_0 = -i\ket{11}_j\ket{1}_0
    \end{align}
    \begin{align}
        \hat{\mathcal{S}}_j\ket{10}_j\ket{0}_0 = -i\ket{11}_j\ket{0}_0 && \hat{\mathcal{S}}_j\ket{10}_j\ket{1}_0 = \ket{01}_j\ket{1}_0
    \end{align}
    Since we adopted the direct mapping, we are not interested in states $\ket{11}_j\ket{0}_0$ or $\ket{11}_j\ket{1}_0$. Application of $\mathcal{U}_m$ leaves the second qubit of each mode in state $\ket{1}$, given that the initial state was $\ket{g_m}$ or $\ket{e_m}$. These states transform as follows:
    \begin{equation}
        \hat{\mathcal{U}}_m\ket{g_m} = (-i)^{m-1}\bigotimes_{j=1}^{m-1}\ket{11}_j\ket{11}_0
        \label{eq:Ustate-g}
    \end{equation}
    \begin{equation}
        \hat{\mathcal{U}}_m\ket{e_m} = (-i)^{m-1}\bigotimes_{j=1}^{m-1}\ket{11}_j\ket{10}_0
        \label{eq:Ustate-e}
    \end{equation}
    As expected, the states given in eqs.~\ref{eq:Ustate-g},\ref{eq:Ustate-e} differ only on the first qubit. Application of the controlled Y-rotation followed by the inverse of $\hat{\mathcal{U}}_m$ retains the transformation given in eqs.~\ref{eq:rot1},\ref{eq:rot2}. Figure~\ref{fig:example} displays an example circuit implementing $\hat{U}_m(\theta)$ for $m=3$.

\section{Cost analysis}
\label{sec:cost}
Table~\ref{tab:exp} compares methods for decomposing $U_m(\theta)$ defined in eq.~\ref{eq:exc_unitary}. CNOT gate counts are given for the exponential method described in ref.~\cite{yordanov2020}, the method of Givens rotations~\cite{yordanov2020,Nam2020,Anselmetti_2021} and our new proposal discussed in sec.~\ref{sec:procedure}.
The exponential method requires $4^m(2m-1)$ CNOT gates to implement the $m$-excitation unitary $U_m(\theta)$. The best to our knowledge literature approach requires $2(2m-1)$ CNOT's and a $(2m-1)$-controlled rotation~\cite{yordanov2020,Nam2020,Anselmetti_2021}, while our solution carries the cost of $(8m-6)$ CNOT's plus a single $m$-controlled rotation. 

The exact CNOT count comparison depends on the choice of method for decomposition of multi-controlled rotations. We consider two approaches. The most straightforward decomposition is described in ref.~\cite{yordanov2020} and requires $2^m$ CNOT gates for a $m$-qubit controlled rotation. Table~\ref{tab:exp} compares CNOT counts for 4 lowest excitations levels assuming this \textit{exponential cost} decomposition of multi-controlled rotations.
\begin{table}[h]
    \begin{tabular}{c|c|c|c}
        Excitation level $m$&Exponential~\cite{yordanov2020}& Givens rotations~\cite{Nam2020,Anselmetti_2021} & This work \\
        \hline
        1 & 4 &3 &3\\
        2& 48 &13 &13\\
        3& 320 &41 &25\\
        4& 1792 &142 &42\\       
    \end{tabular}
	\captionsetup{justification=raggedright, singlelinecheck=false}
\caption{Comparison of the CNOT gate counts for implementations of the $m$-excitation unitary from eq.~\ref{eq:exc_unitary} for $m=1,2,3,4$. Columns represent respectively: excitation level $m$, CNOT gate count for: the exponential decomposition method~\cite{yordanov2020}, method based on Givens rotations~\cite{Nam2020,Anselmetti_2021}, our proposed method. All methods assume an identical scheme for decomposing a $n$-controlled rotation into $2^n$ CNOT gates.}
    \label{tab:exp}
\end{table}

The other approach uses ancilla qubits for a CNOT decomposition of the multi-controlled rotation, which scales linearly in $m$. Instead focusing on a particular approach, we consider a class of methods that implements C$^nX$ with $(An-B)$ Toffoli gates, where $A$, and $B$ are integer coefficients. Details of such methods can be found in \cite{Nielsen_Chuang_2010,ElementGates,khattar2024riseconditionallycleanancillae,Silva2023LinearDO}. 
For example, in ref.~\cite{ElementGates}, the $n$-controlled rotation is implemented with two $C^{n-1}X$ gates and two single-controlled rotations. The $C^{n-1}X$ gate can be further decomposed with ancilla qubits using a linear number of Toffoli gates (CNOT gates).

Implementations of $U_m(\theta)$ with Givens rotation, as given in  eqs.~\ref{eq:vib_U}, can use relative Toffoli gates instead of the complete Toffoli gates. Relative Toffoli gates can be decomposed into three CNOT gates and four T-gates (for fault-tolerant architectures), compared to six two-qubit gates and seven T-gates for the complete Toffoli gate. 
Table~\ref{tab:CnotsLin} compares the CNOT counts between the literature methods based on Givens rotations~\cite{Nam2020,Anselmetti_2021} and our technique, adopting an ancilla-based C$^nX$ to $(An-B)$ Toffoli gate decomposition. 

The second column in ~\ref{tab:CnotsLin} gives a general expression for the number of CNOT gates required to implement $U_m(\theta)$ with the Toffoli gates. The third column gives the CNOT count when the relative Toffoli gates are used. Our proposed method is advantageous in the CNOT count even for $A=1$ and the advantage reaches 50\% in the limit of large $A$. These conclusions remain true for decompositions of $C^nX$ with the relative phase Toffoli gate. The fourth column in Table~\ref{tab:CnotsLin} gives the CNOT count for a particular method proposed in \cite{khattar2024riseconditionallycleanancillae}, with  $A =2$, $B=3$. In this  case, our method gives a $28\%$ reduction in CNOT count with respect to ref.~ \cite{khattar2024riseconditionallycleanancillae}.

\begin{table}[h]
    \begin{tabular}{c|c|c| c}
    Method & TOF & RTOF & RTOF (A=2, B=3)~\cite{khattar2024riseconditionallycleanancillae}\\
    \hline
    Givens~\cite{Arrazola2022,Nam2020,Anselmetti_2021} & (24A+4)m-24A-12B+2 & (12A+4)m-12A-6B+2& 28m-40\\
    This work &(12A+8)m-12A-12B-2 & (6A+8)m-6A-6B-2& 20m-32         
    \end{tabular}
	\captionsetup{justification=raggedright, singlelinecheck=false}
    \caption{Comparison of the CNOT gate counts for implementations of the $m$-excitation unitary from eq.~\ref{eq:exc_unitary}. Columns represent respectively: decomposition method, CNOT gate count for: ancilla-assisted C$^nX$ to $(An-B)$ Toffoli gate decomposition and full Toffoli gate decomposition into CNOTs, ancilla-assisted C$^nX$ to $(An-B)$ Toffoli gate decomposition and relative Toffoli gate decomposition into CNOTs, ancilla-assisted C$^nX$ to $(An-B)$ Toffoli gate decomposition and relative Toffoli gate decomposition into CNOTs for A=2 and B=3 of ref.~\cite{khattar2024riseconditionallycleanancillae}}
    \label{tab:CnotsLin}
\end{table}

Our technique requires half the control qubits of standard Givens-rotation-based methods~\cite{khattar2024riseconditionallycleanancillae,Silva2023LinearDO,Nam2020,Anselmetti_2021}. Consequently, for $C^nX$ decompositions, for which the number of ancilla qubits scales as $\mathcal{O}(n)$, the cost of implementing $n$-controlled rotations in our method is reduced by half compared to the Givens rotation approach. An extra Toffoli gate count overhead in our method appears in the $\hat{\mathcal{U}}$ operations given in eq.~\ref{transf_basis}, which scales as $\mathcal{O}(m)$.
Overall, in fault-tolerant implementations, where Toffoli gate count is a critical resource, our method for implementing eq.~\ref{eq:U_decomp} yields a Toffoli count of $2(A+2)m+C$ compared to $4Am+C'$ for the Givens rotations method.

\newpage
\section{Demonstration on quantum hardware}
\label{sec:exp}

The performance of our UVCC state preparation method was validated experimentally by measuring circuit execution fidelity on quantum hardware. We compared our approach with the Givens rotations method~\cite{Arrazola2022, Nam2020, Anselmetti_2021}, as discussed in sec.~\ref{sec:UVCC}. For these experiments, we selected a test system with three non-degenerate vibrational modes, labeled $M_0$, $M_1$, and $M_2$. 
\begin{figure}[!h]
	\subfloat[System \textit{S-6}: This work. TVD = 0.064]{\includegraphics[width=3in]{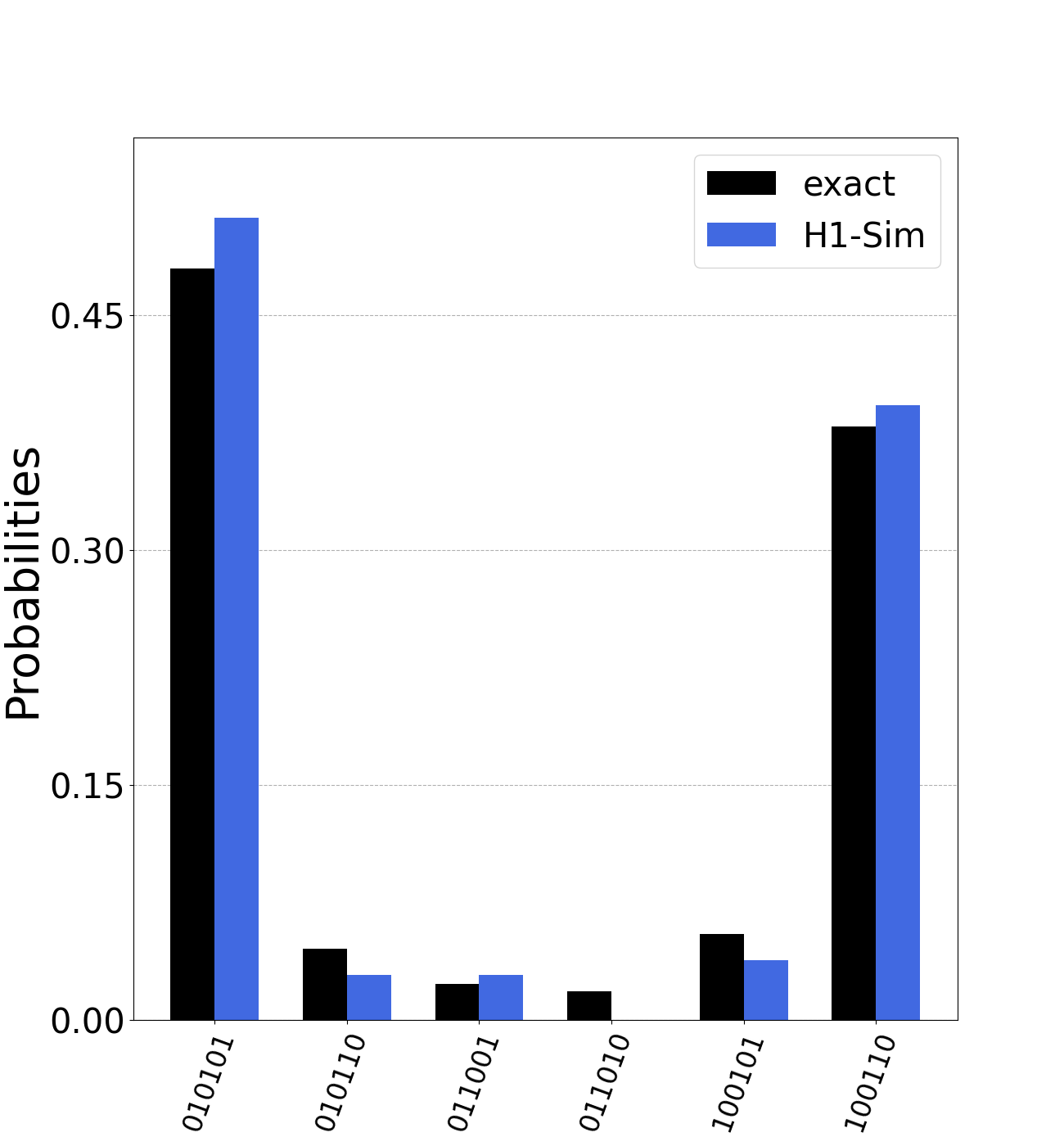}}
	\subfloat[System \textit{S-6}: Givens rotations. TVD = 0.082]{\includegraphics[width=3in]{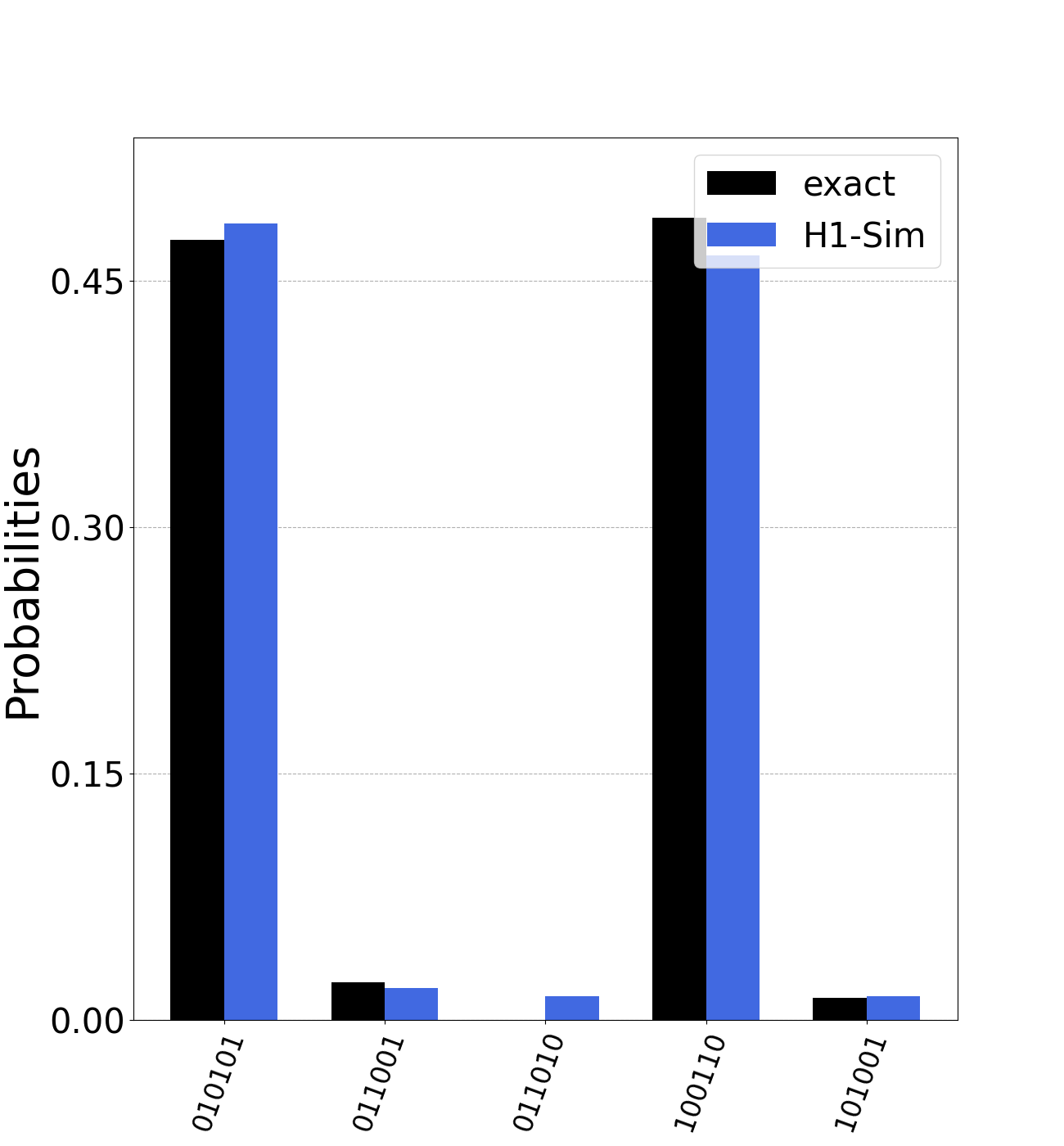}}\\
	\subfloat[System \textit{S-8}: This work. TVD = 0.242]{\includegraphics[width=3in]{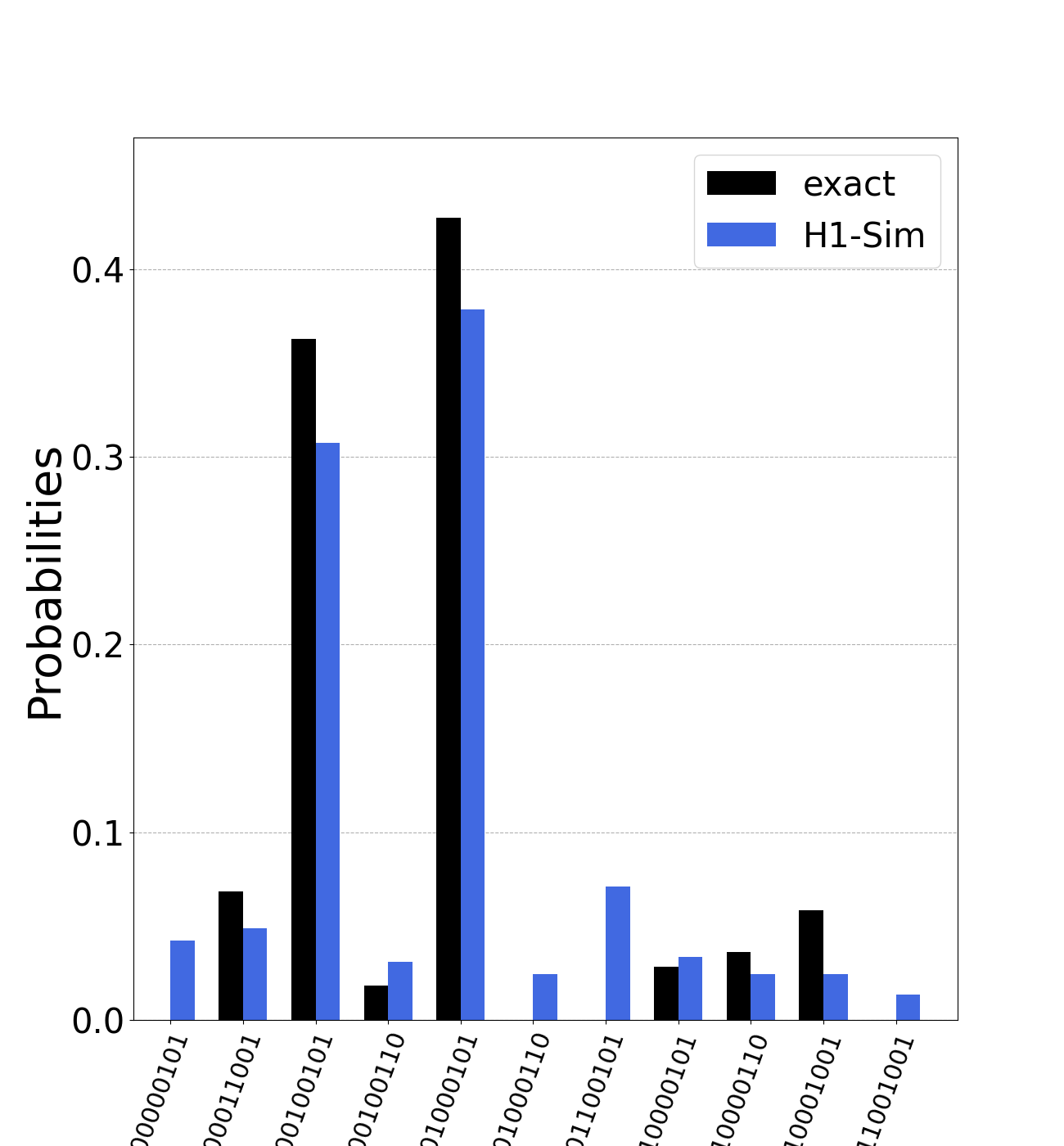}}
	\subfloat[System \textit{S-8}: Givens rotations. TVD = 0.469]{\includegraphics[width=3in]{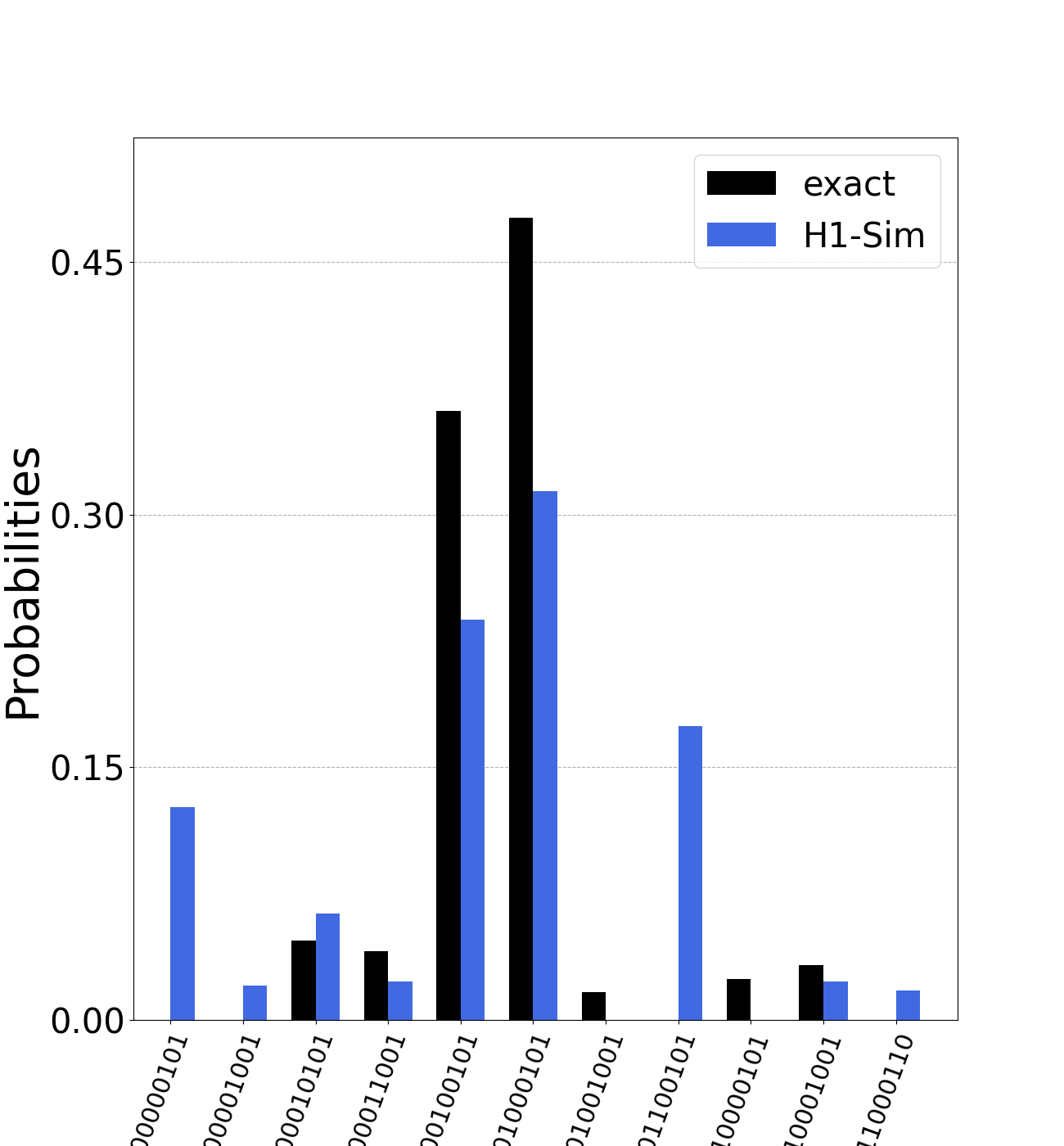}}
	\captionsetup{justification=raggedright, singlelinecheck=false}
	\caption{Histograms comparing state probability distributions obtained by running the UVCCSDT ansatz on Quantinuum's H1-1 Emulator (\textit{H1-Sim}) and noiseless Qiskit Simulator (\textit{exact}). a) System \textit{S-6} implemented with the present method; b)  System \textit{S-6} implemented with the Givens rotation method; c) System \textit{S-8} implemented with the present method; d) System \textit{S-8} implemented with the Givens rotations method. For each experiment the total variation distance (TVD) was calculated. TVD is defined as: $TVD(P,Q) = \sup_{A \in \mathcal{A}}|P(A)-Q(A)|$, where $P$ and $Q$ are probability measures and $\mathcal{A}$ is an appropriate $\sigma$-algebra \cite{Tsybakov2009}.} 
	\label{fig:results_sim}
\end{figure}
Two model systems, denoted \textit{S-6} and \textit{S-8}, were considered.
System \textit{S-6} includes two basis states per mode, encoded into six qubits, while system \textit{S-8} has two basis states for modes $M_0$ and $M_1$ and four states for mode $M_2$, requiring eight qubits. In both cases, the reference state is the ground state $\ket{0,0,0}$.

State preparation circuits for systems \textit{S-6} and \textit{S-8} implement the UVCC ansatz given in eq.~\ref{eq:UVCC_ansatz} with full single-, double- and triple-excitations, $m=1,2,3$, and are denoted as UVCCSDT.
The angles that parametrize the UVCC ansatz were chosen arbitrarily to cover the excitation space with amplitudes distributed among several states, as detailed in Appendix~\ref{sec:appendix}. 
\begin{figure}[!h]
	\subfloat[System \textit{S-6}: This work. TVD = 0.050]{\includegraphics[width=3in]{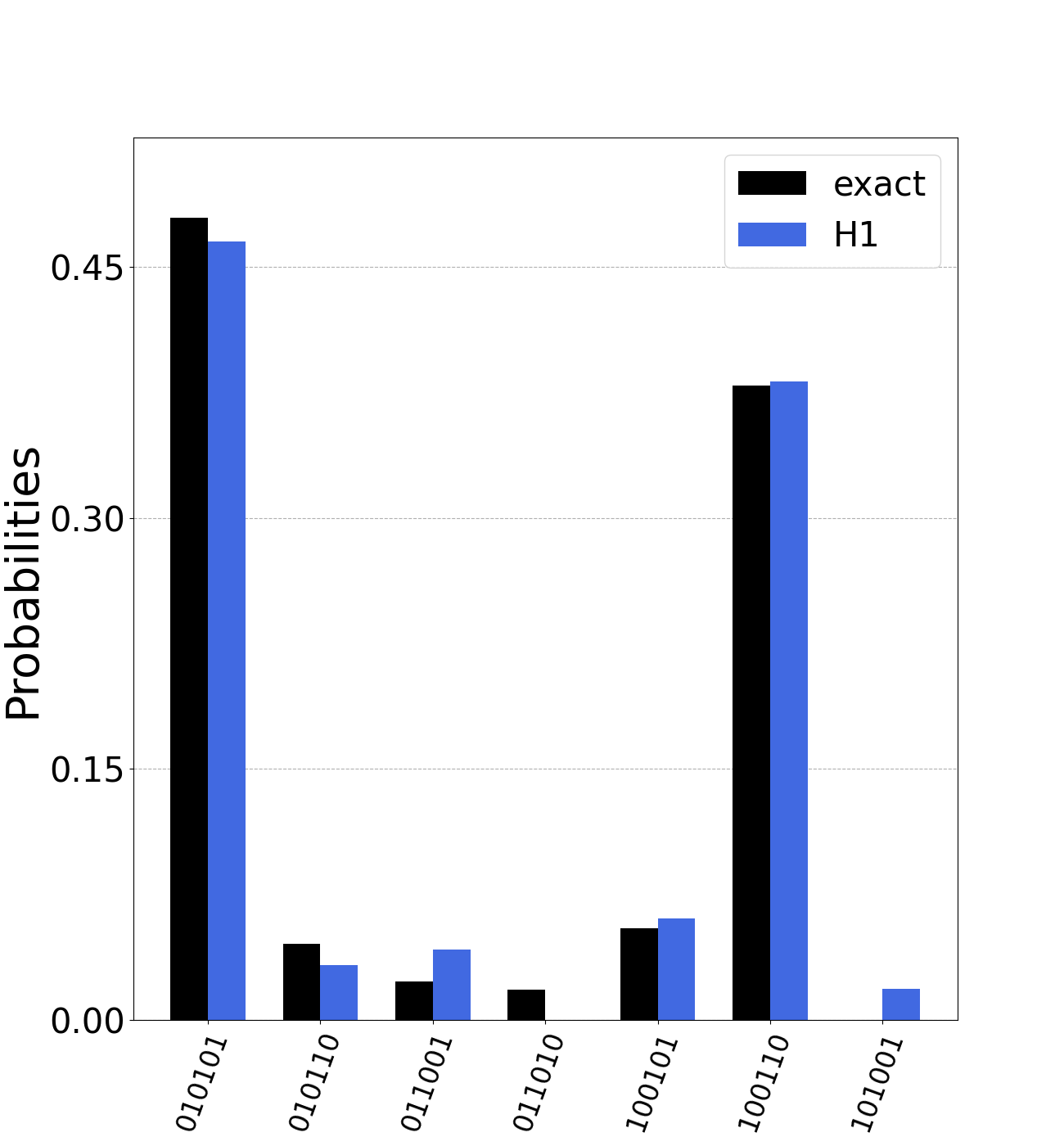}}
	\subfloat[System \textit{S-6}: Givens rotations. TVD = 0.136]{\includegraphics[width=3in]{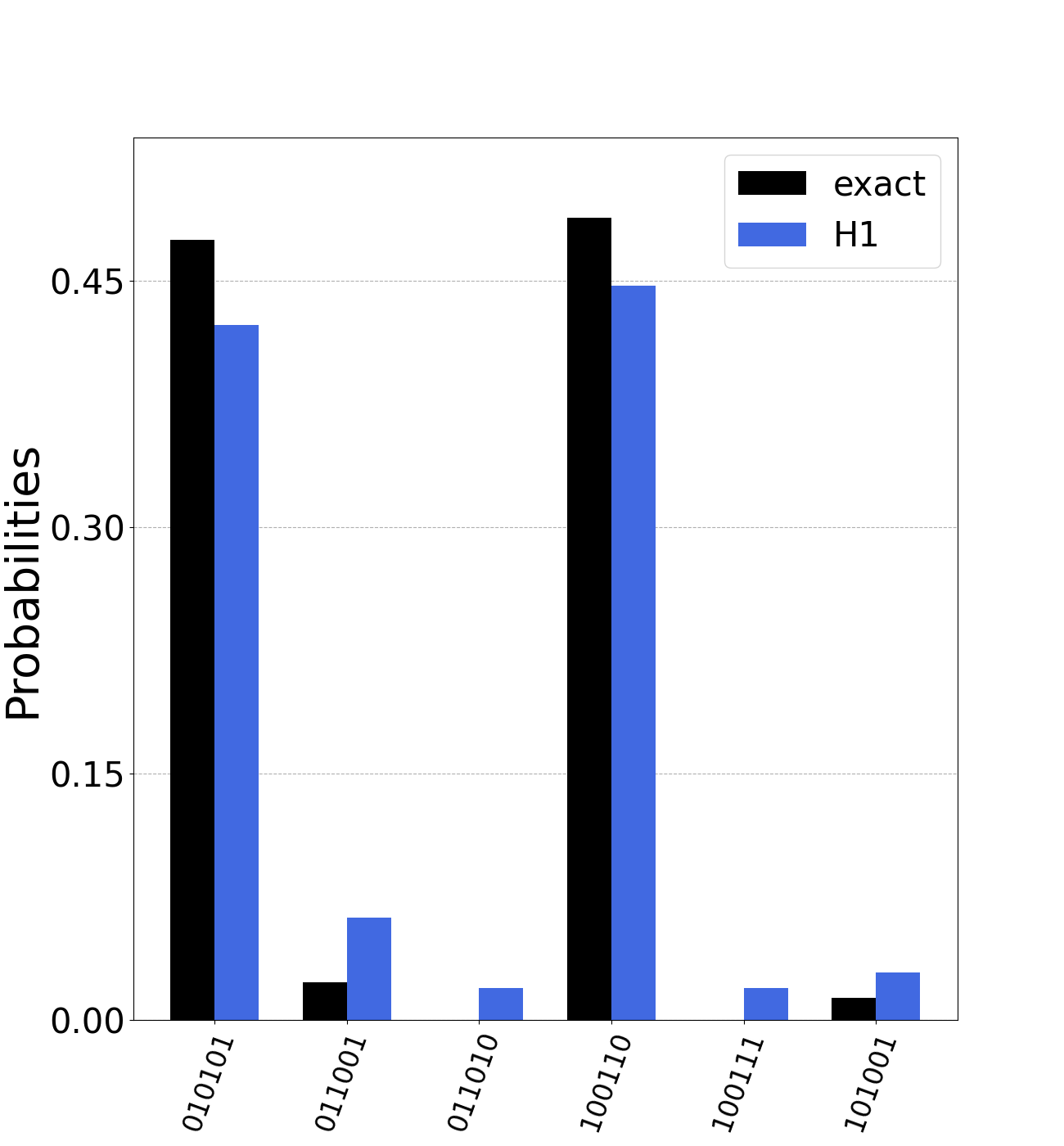}}\\
	\subfloat[System \textit{S-8}: This work. TVD = 0.240]{\includegraphics[width=3in]{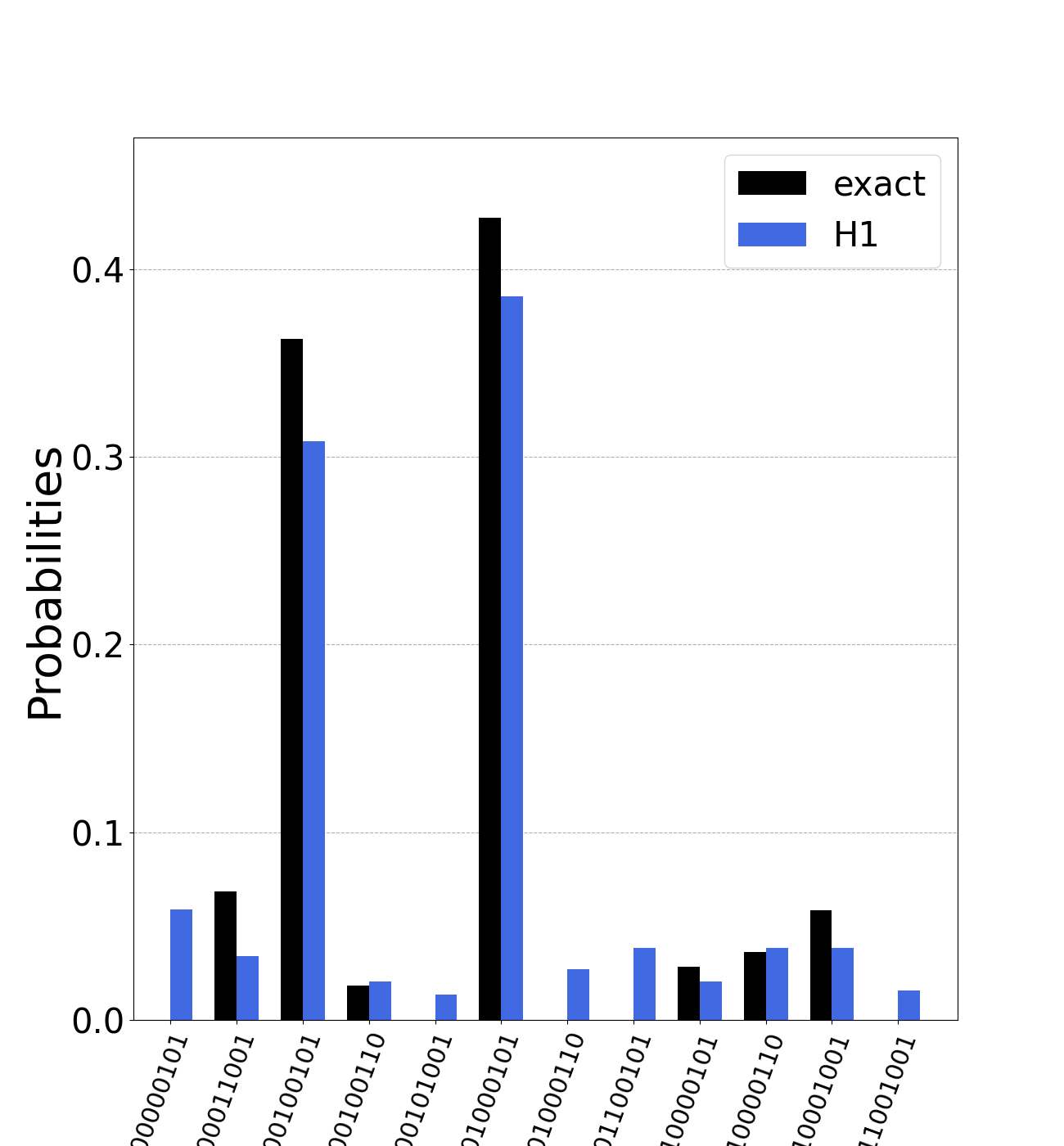}}
	\subfloat[System \textit{S-8}: Givens rotations. TVD = 0.320]{\includegraphics[width=3in]{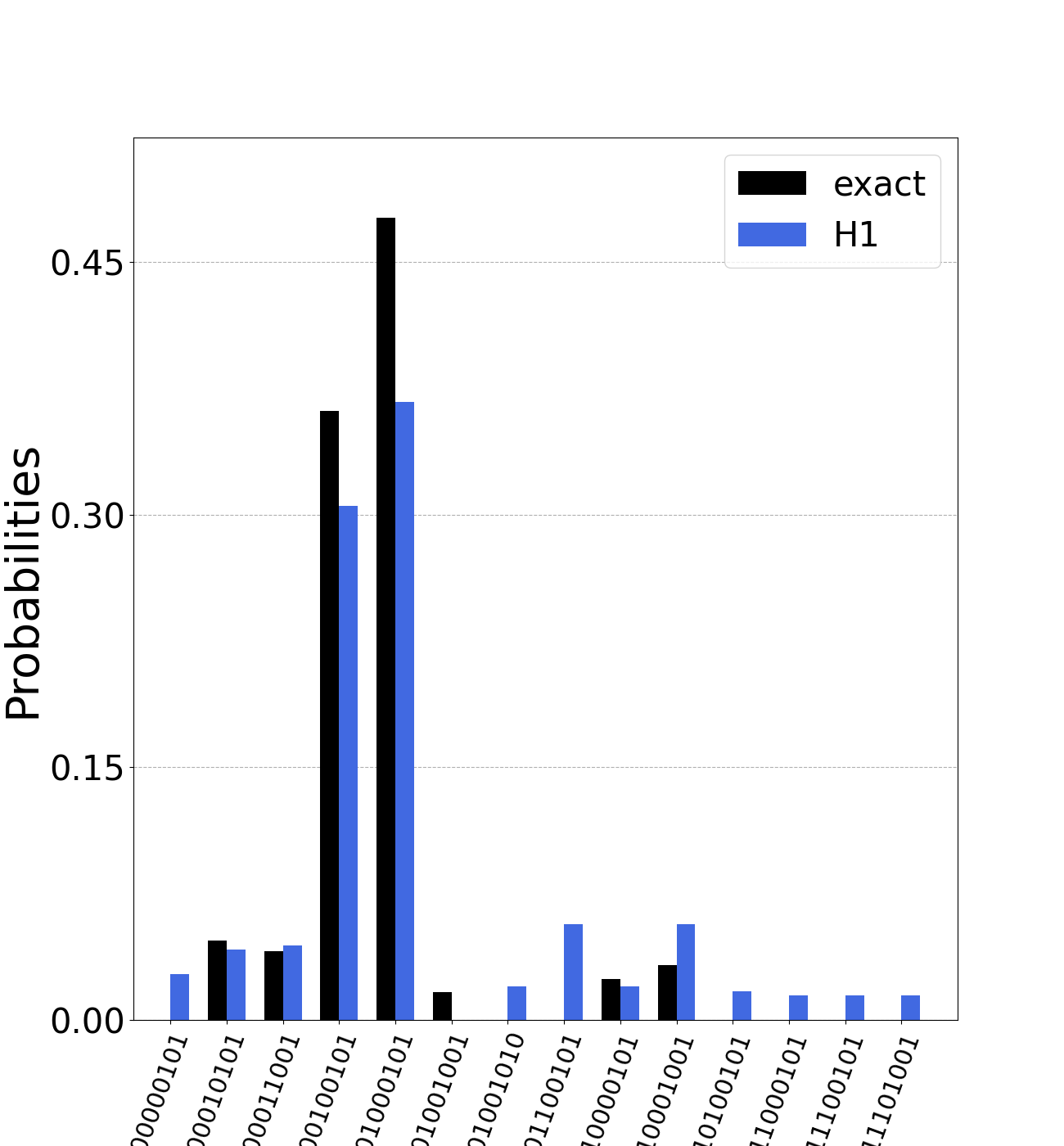}}
	\captionsetup{justification=raggedright, singlelinecheck=false}
	\caption{Histograms comparing state probability distributions obtained by running the UVCCSDT ansatz on Quantinuum's H1-1 quantum device (\textit{H1}) and noiseless Qiskit Simulator (\textit{exact}). a) System \textit{S-6} implemented with the present method; b)  System \textit{S-6} implemented with the Givens rotation method; c) System \textit{S-8} implemented with the present method; d) System \textit{S-8} implemented with the Givens rotations method. For each experiment the total variation distance (TVD) was calculated. Systems \textit{S-6} and \textit{S-8} were sampled with 220 and 512 shots, respectively.}
	\label{fig:results_hardware}
\end{figure}
Since the UVCC operator prepares the target state by acting on qubits initialized in $\ket{0}$ states, we optimized both our method and the Givens rotations approach by removing unnecessary entangling gates at the start of the circuit. 
We performed consistent optimizations of quantum circuits for our procedure and for the method of Givens rotations. 
Before running on hardware, we simulated the quantum circuits using Quantinuum’s H1-1 emulator, with results shown in Fig.~\ref{fig:results_sim}.

\begin{figure}[!h]
	\centering
	\includegraphics[width=1\textwidth]{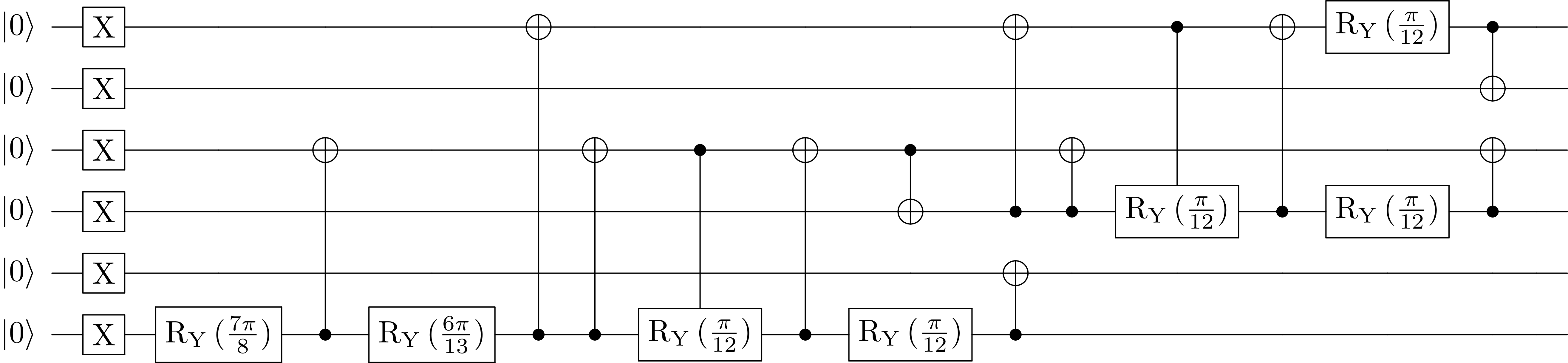}
		\captionsetup{justification=raggedright, singlelinecheck=false}
	\caption{Quantum circuit implementing triple-excitation UVCC ansatz (UVCCSDT) for system \textit{S-6} with the method described in sec.~\ref{sec:procedure}. Qubits are numbered 0 to 5, top to bottom. The state of mode $M_j$ is mapped to qubits $(q_{2j},q_{2j+1})$ with $j = 0,1,2$. The number of controlled gates was further reduced by removing unnecessary controls on qubits in known initial state.}
	\label{fig:circuitA}
\end{figure}

\begin{figure}[!h]
	\centering
	\includegraphics[width=1\textwidth]{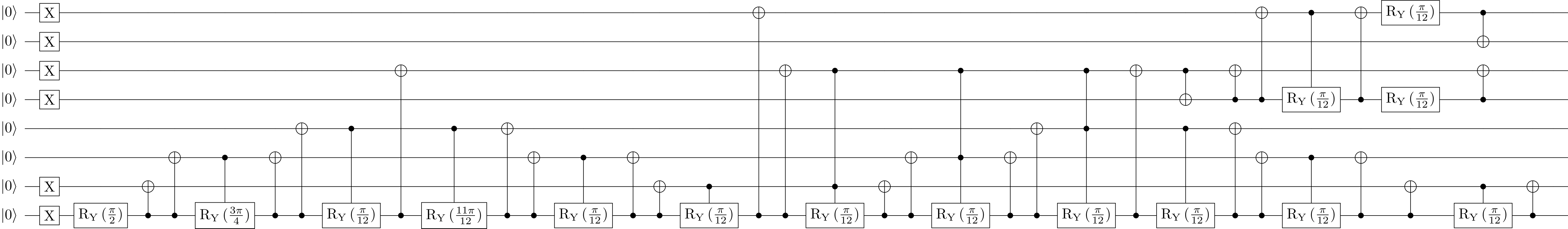}
	\caption{
		Quantum circuit implementing triple-excitation UVCC ansatz (UVCCSDT) for system \textit{S-8} with the method described in sec.~\ref{sec:procedure}. Qubits are numbered 0 to 7, top to bottom. The states of modes $M_0$, $M_1$ and $M_2$ are mapped to qubits $(q_0,q_1), (q_2,q_3)$ and $(q_4,q_5,q_6,q_7)$ respectively. The number of controlled gates was further reduced by removing unnecessary controls on qubits in known initial state.}
	\label{fig:circuitB}
\end{figure}

Fig.~\ref{fig:results_hardware} shows results of experiments with the H-1 quantum machine for systems \textit{S-6} and \textit{S-8}, with 220 and 512 shots, respectively. Circuits implementing our method for systems \textit{S-6} and \textit{S-8} are given in Figs.~\ref{fig:circuitA} and \ref{fig:circuitB}, respectively.

The corresponding circuits for the Givens rotation method are given in Appendix~\ref{sec:appendix}.
Both emulator results and quantum hardware experiments confirm that the circuits developed in the present work achieve higher fidelities in UVCC state preparation than the Givens rotation method. For system \textit{S-6}, our circuit achieves a Total Variation Distance (TVD) of 0.050, compared to 0.136 for the Givens rotation method. For the 8-qubit system \textit{S-8}, the TVD for our circuit is 0.240, while the Givens rotation method yields a TVD of 0.320.

\section{Summary}
We proposed a new method for implementing the Unitary Vibrational Coupled-Cluster (UVCC) ansatz on a quantum computer. By utilizing redundancies in the Hilbert space for the direct qubit mapping, our approach achieves a theoretical reduction of up to $50\%$ in the number of entangling CNOT gates compared to existing techniques. When compared to practically useful approaches, our procedure offers up to a $28\%$ reduction in CNOT count, with the maximum benefit observed at high vibrational excitations in the UVCC ansatz.
The precise advantage of our method depends on the decomposition of multi-controlled arbitrary rotations into Toffoli gates. Notably, our implementation of multi-controlled rotations reduces the number of qubit controls in the Trotterized UVCC ansatz by half, lowering gate counts for both near-term intermediate-scale quantum (NISQ) devices and, in specific cases, for Fault-Tolerant quantum architectures. We validated the performance of our new method on Quantinnum's H1-1 quantum hardware and emulators, observing fidelity improvements in experiments with 6- and 8-qubit systems. For the 6 qubit system, our method improved the total variation distance metric by 172\%, while for the 8-qubit system, it yielded a 33\% improvement. 
These gains stem from the reduced entangling gate count in our method. Our technique thus carries the potential to enable more complex vibrational structure calculations on quantum computers earlier.

\section{Acknowledgments}
This work is funded by the European Innovation Council accelerator grant COMFTQUA, no. 190183782.

\section*{Appendix A}
\label{sec:appendix}
Details for the experimental implementations of the UVCC ansatz are discussed in sec.~\ref{sec:exp}.
For the UVCCSDT ansatz applied to system \textit{S-6}, the chosen angles are: $\pi[\frac{7}{8},\frac{6}{13}, \frac{1}{12},\frac{1}{12},\frac{1}{12},\frac{1}{12},\frac{1}{12}]$ corresponding to the states $\ket{111},\ket{101},\ket{110},\ket{011},\ket{100},\ket{010},\ket{001}$, respectively.
For system \textit{S-8} the angles are given by $\frac{\pi}{12}[6,9,1,11,1,1,1,1,1,1,1,1,1,1,1]$ and correspond to states:
$[\ket{1,1,1},\ket{2,1,1},\ket{3,1,1},\ket{3,0,1},\ket{2,0,1}$,
$\ket{1,0,1},\ket{1,1,0},\ket{2,1,0},\ket{3,1,0},\ket{0,1,1},\ket{0,0,1},\ket{0,1,0},\ket{3,0,0},\ket{2,0,0},\ket{1,0,0}]$, respectively.

\begin{figure}[H]
	\centering
	\includegraphics[width=1\textwidth]{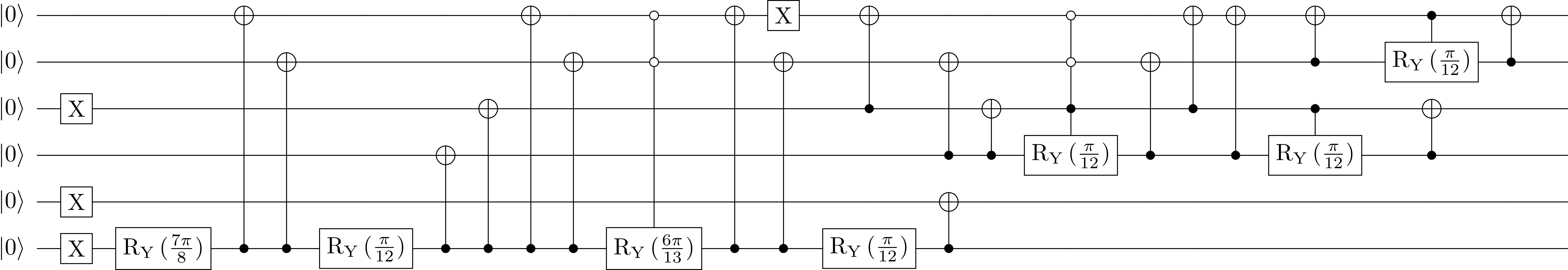}
	\caption{
		Quantum circuit implementing the triple-excitation UVCC ansatz (UVCCSDT) for system \textit{S-6} with the method described in refs.~\cite{Nam2020,Anselmetti_2021}. Qubits are numbered 0 to 5, top to bottom. The state of mode $M_j$ is mapped to qubits $(q_{2j},q_{2j+1})$ with $j = 0,1,2$. The number of controlled gates was further reduced by removing unnecessary controls on qubits in known initial state.}
	\label{fig:circuitA_Givens}
\end{figure}
\begin{figure}[H]
	\centering
	\includegraphics[width=1\textwidth]{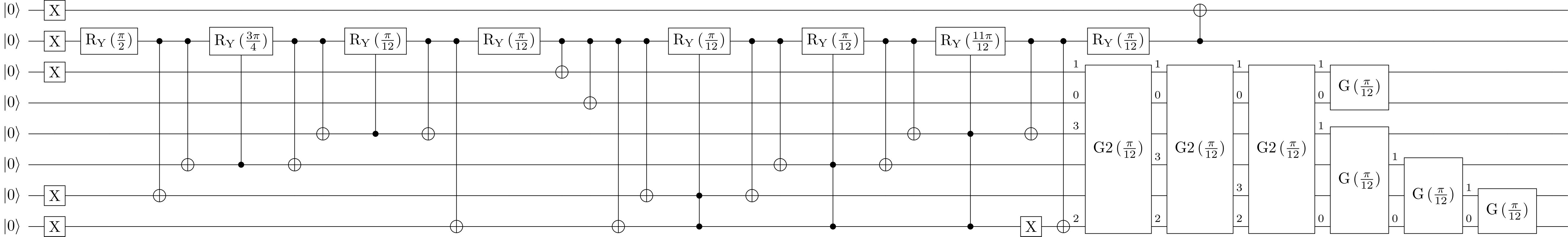}
	\caption{
		Quantum circuit implementing the triple-excitation UVCC ansatz (UVCCSDT) for system \textit{S-8} with the method described in refs.~\cite{Nam2020,Anselmetti_2021}. Qubits are numbered 0 to 7, top to bottom. The states of modes $M_0$, $M_1$ and $M_2$ are mapped to qubits $(q_0,q_1), (q_2,q_3)$ and $(q_4,q_5,q_6,q_7)$ respectively. The number of controlled gates was further reduced by removing unnecessary controls on qubits in known initial state. For clarity, decomposition of the excitation unitary is shown explicitly for the part of the circuit that has undergone a systematic optimization, consistent with our method. The remaining gates are depicted as a general single and double excitation $G$ and $G2$~\cite{Arrazola2022} shown in Fig.~ \ref{fig:G} and Fig.~\ref{fig:G2}, respectively.}
	\label{fig:circuitB_Givens}
\end{figure}
\begin{figure}[H]
	\centering
	\includegraphics[width=0.4\textwidth]{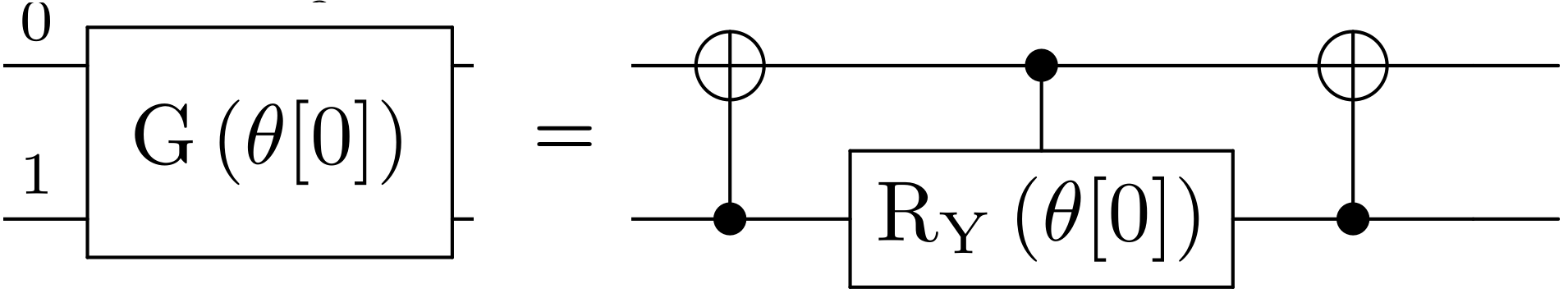}
	\caption{Single excitation unitary~\cite{Arrazola2022} implemented with a method from ref.~\cite{Nam2020,Anselmetti_2021}.}
	\label{fig:G}
\end{figure}
\begin{figure}[H]
	\centering
	\includegraphics[width=0.4\textwidth]{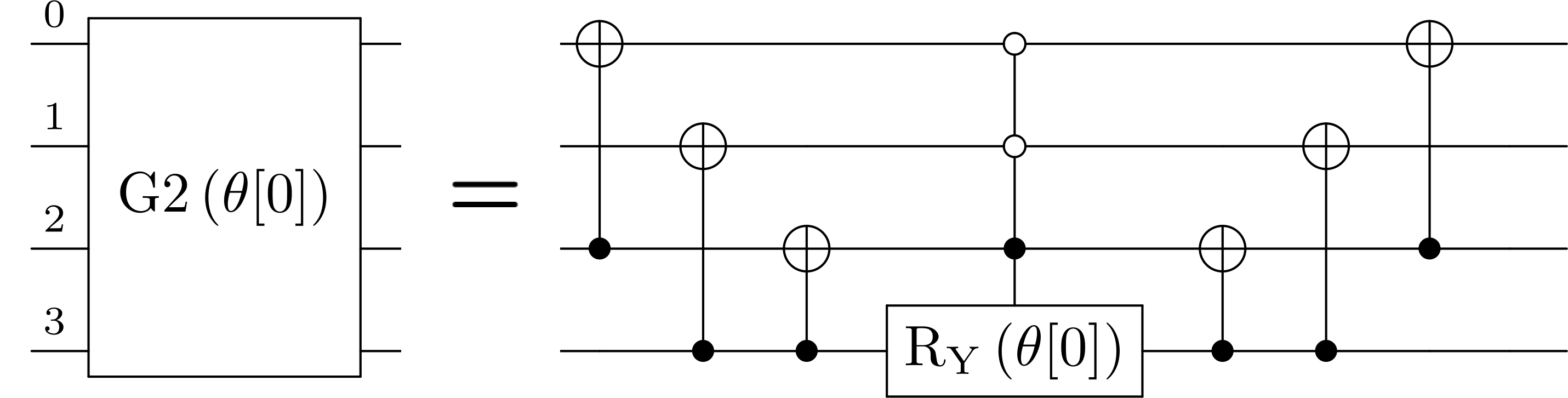}
	\caption{Double excitation unitary~\cite{Arrazola2022} implemented with a method from ref.~\cite{Nam2020,Anselmetti_2021}.}
	\label{fig:G2}
\end{figure}

\bibliography{sources.bib}

\end{document}